\documentstyle[12pt]{article}
\begin{document}
\begin{center}
\renewcommand{\thefootnote}{\fnsymbol{footnote}}

{\bf \Large Parton Sum Rules and Nucleon Structure Functions
\footnote{Published in Proceedings of the Second International
Symposium on Medium Energy Physics, Beijing, China, August 22-26,
1994, Edited by Weiqin Chao \& Pengnian Shen, World Scientific,
Singapore, 1995,
p.298-300}}

\vspace{3mm}
{\bf \large Bo-Qiang Ma
}
\vspace{2mm}

{\large 
Institute of High Energy Physics, Academia Sinica, P.O.Box 918(4),
Beijing 100039, China}
\vspace{5mm}

\small Abstract
\end{center}

\footnotesize

A brief summary on the Wigner rotation effect in the understanding
of the proton spin ``crisis" related with the Ellis-Jaffe sum rule 
violation, and on the proton-neutron isospin symmetry breaking 
explanation of the Gottfried sum rule violation. 
The proton spin could be fully provided by the spin sum of quarks 
if one considers the Wigner rotation effect in the measured 
quark helicity distributions.

\vspace{5mm}

\small

Parton sum rules provide information of the quark distributions in
the nucleons and are very useful to reveal new physics if a sum rule
is found to be satisfied or broken.
Parton sum rules or similar relations played important roles in
the establishment of the quark-parton model. 
Recently there have been 
new interests in the Ellis-Jaffe sum rules (EJSR) and
Gottfried sum rule (GSR). 
They are found to be broken by the EMC and NMC experiments. 
This inspired a vast number of studies on the explicit spin 
and flavor distributions in the nucleons.  

\noindent
{\bf 1. The violation of the Ellis-Jaffe sum rules 
and the proton spin
``crisis''}

The EMC result of a smaller integrated spin-dependent structure 
function data than that expected from the Ellis-Jaffe sum rule 
triggered the proton ``spin crisis'', i.e., the
intriguing question of 
how the spin of the proton is distributed among
its quark spin, gluon spin and orbital angular momentum. At present
it is commonly taken for granted that the EMC result implies that
there must be some contribution due to gluon polarization or orbital
angular momentum to the proton spin. For example, in the gluonic 
and the strange sea explanations of the EJSR breaking, the proton
spin carried by the spin of quarks was estimated to be of about
$70\%$
in the former and negligible in the latter. We have shown
\cite{Spin1,Spin2},
however, that the above acquaintances  are not in contradiction 
with the naive quark model in which the spin of the proton, 
when viewed in its
rest reference frame, is fully provided by the vector sum of the
quark spin. 

The key points for understanding the proton spin puzzle lie in the
fact that the vector sum of the constituent  spin for a composite
system is not Lorentz invariant by taking into account the
relativistic effect of Wigner rotation, and that it is in the
infinite momentum frame the small EMC result was interpreted as
an indication that quarks carry a small amount of the total spin of
the proton. 
From the first fact we know that the vector spin structure of
hadrons could be quite different in different frames from
relativistic viewpoint. We thus can naturally understand the proton
``spin crisis'' because there is no need to require that the sum of 
the quark spin
be equal to the spin of the proton in the infinite
momentum frame, even if the vector sum of the quark spin equals to the
proton spin in the rest frame. 
It is necessary to clarify what is meant by the quantity
$\Delta q$ defined by
$\Delta q\!\cdot\!S_{\mu}
=<\!P,S|\bar{q}\gamma_{\mu}\gamma_{5}q|P,S\!>$, 
where
$S_{\mu}$ is the proton polarization vector. $\Delta q$ can be
calculated from $\Delta q=<\!P,S|\bar{q}\gamma^{+}\gamma_{5}q|P,S\!>$
since the instantaneous fermion lines do not contribute to the +
component. One can easily prove, by expressing the quark wave
functions in terms of light-cone Dirac spinors (i.e., the quark
spin states in the infinite momentum frame), that 
\begin{equation}
\Delta q=\int_{0}^{1}dx\:[q^{\uparrow}(x)-q^{\downarrow}(x)],
\end{equation}
where $q^{\uparrow}(x)$ and $q^{\downarrow}(x)$ are the probabilities
of finding, in the proton infinite momentum frame, a quark or
antiquark of flavor q with fraction x of the proton longitudinal
momentum and with polarization parallel or antiparallel to the proton
spin, respectively. However, if one expresses the quark wave
functions in terms of conventional instant form Dirac spinors
(i.e., the quark spin state in the proton rest frame), it can be
found, that 
\begin{equation}
\Delta q=\int\!d^{3}\vec{p}\:M_{q}\:[q^{\uparrow}(p)-q^{\downarrow}(p)]
=<\!M_{q}\!>\Delta q_{L},
\end{equation}
with 
\begin{equation}
M_{q}=[(p_{0}+p_{3}+m)^{2}-\vec{p}_{\perp}^{2}]/[2(p_{0}+p_{3})
(m+p_{0})]
\end{equation}
being the contribution from the relativistic effect due to the quark
transversal motions, 
$q^{\uparrow}(p)$ and $q^{\downarrow}(p)$ being the
probabilities of finding, in the proton rest frame, a quark or
antiquark of flavor $q$ with rest mass $m$ and momentum $p_{\mu}$ and
with spin parallel or antiparallel to the proton spin
respectively, and $\Delta q_{L}=\int\!d^{3}\vec{p}
[q^{\uparrow}(p)-q^{\downarrow}(p)]$ being the net spin vector sum of
quark flavor $q$ parallel to the proton spin in the rest frame. Thus
one sees that the quantity $\Delta q$ is better to
be interpreted as the
net spin polarization in the infinite momentum frame if one properly
considers the relativistic effect due to internal quark transversal
motions.

In both the gluonic and strange sea explanations 
of the EJSR violation,
the spin of the proton could be fully provided by the vector sum of 
the quark spin in the naive quark model 
based on our above
definition of the spin structure of a composite system and above
clarification of the physical implication of the quantity $\Delta q$.
Since $<\!\!M_{q}\!\!>$, the average contribution from the relativistic
effect due to internal transversal motions of quark flavor $q$,
ranges from $0$ to $1$, and $\Delta q_{L}$, 
the net spin vector polarization
of quark flavor $q$ parallel to the proton spin in the proton rest
frame, is related to the quantity $\Delta q$ by the relation 
$\Delta q_{L}=\Delta q/<\!\!M_{q}\!\!>$, 
we have sufficient freedom to make
the naive quark model spin sum rule, 
i.e., 
$\Delta u_{L}+\Delta d_{L}+\Delta s_{L}=1$, 
satisfied while still preserving 
the values of $\Delta u$,
$\Delta d$ and $\Delta s$ required
in the two
explanations, respectively. In the gluonic explanation, we could
choose $\Delta u_{L}=4/3$, $\Delta d_{L}=-1/3$, and $\Delta s_{L}=0$
as those in the most simply SU(6) configuration of naive quark model,
then one sees that we need 
$<\!\!M_{u}\!\!>\approx <\!\!M_{d}\!\!> \approx 0.7$ to
preserve 
$\Delta u$ and $\Delta d$ in this explanation.
In the strange sea explanation, the non-vanishing
$\Delta s$ reflects the polarizations of sea thus some number of sea
quarks, then one has large freedom to choose arbitrary 
$\Delta u_{L}$, $\Delta d_{L}$, and $\Delta s_{L}$ constrained by the
naive quark model spin sum rule 
while still preserving 
the values of $\Delta u$,
$\Delta d$ and $\Delta s$ 
in the two
explanations.
In both the 
above cases the proton spin is fully provided by the spin vector sum of
quarks. 
Thereby we can understand 
the ``spin crisis'' simply because the quantity
$\Delta\Sigma=\Delta u+\Delta d+\Delta s$ does not, in a strict
sense, represent the vector sum of the spin carried
by the quarks in the naive quark model.
It is possible that the value of
$\Delta\Sigma=\Delta u+\Delta d+\Delta s$ is small whereas the
spin sum rule
\begin{equation}
\Delta u_{L}+\Delta d_{L}+\Delta s_{L}=1 
\end{equation}
for the naive quark model still holds, 
though the realistic situation may be complicated.
  
\noindent
{\bf 2. The Gottfried sum rule (GSR) violation and sea quark content of
nucleons} 

The measurement of the Gottfried sum
by the NMC collaboration 
inspired a number of investigations on the flavour distributions
in the sea of
nucleons.
In some works this
GSR violation was interpreted as an indication for a flavor
asymmetry of the sea quark in the nucleons, 
but this explanation was obscurely named ``isospin violation''
in some literature.  
It has been observed by us \cite{Flavor1,Flavor2} 
that there is an  alternative possibility that the isospin
symmetry between the proton and the neutron is broken, while still
preserving the flavor symmetry of the sea in the proton and the
neutron. We have 
examined systematically the consequences of this possibility 
for several processes, namely, neutrino deep inelastic
scattering,
charged pion Drell-Yan process, proton Drell-Yan process, and
semi-inclusive 
deep inelastic
scattering. From our investigations we conclude that the two
alternative explanations of the GSR violation, namely, the asymmetric
SU(2) flavor sea proposed by others and the p-n isospin symmetry
violation proposed by us, should give different results for some of
the above processes, and that a decision between the two alternative
explanations is possible. 

Deep inelastic neutrino and anti-neutrino scattering 
on protons and deuterons  provides a good way 
to distinguish between a SU(2) sea asymmetry
and p-n isospin symmetry breaking.
If the SU(2) sea is asymmetry
the violation of GSR indicates
an excess of $d\overline{d}$ over $u\overline{u}$; 
i.e.,
\begin{equation}
\int_{0}^{1}[\overline{u}(x)-\overline{d}(x)]dx
=-0.140\pm0.024.         
\label{eq:su2sea}
\end{equation}
Whereas for p-n isospin symmetry breaking the
violation of GSR is due to an excess of sea quarks in
neutrons over those in protons while preserving the SU(2) symmetry
in the sea of nucleons; i.e.,
\begin{equation}
\int_{0}^{1}[\overline{q}^{p}(x)-\overline{q}^{n}(x)]dx
=-0.084\pm0.014.        
\label{eq:pnsea}
\end{equation}
\noindent
Both explanations are fitted to the observed
$S_{G}$ but 
may give very different values for some linear combinations of
neutrino structure functions from protons and deuterons. 
We suggested
a new sum, defined by
\begin{equation}
S=\int_{0}^{1}{[(F_{2}^{\nu p}+F_{2}^{\overline{\nu}p})-
\frac{1}{2}(F_{2}^{\nu D}+F_{2}^{\overline{\nu} D})]}dx.
\label{eq:ns}
\end{equation}
This new sum is zero for an asymmetric sea 
explanation and 
 $4\int_{0}^{1}[\overline{q}^{p}(x)-\overline{q}^{n}(x)]dx=
-0.336\pm0.058$ for p-n symmetry breaking, thus well suited to
distinguish
between the two alternative explanations.

We also examied the Drell-Yan processes in both charged pion
and proton scatterings on nuclear targets \cite{Flavor2,Flavor3}. 
We found that the
two explanations give approximately the same predictions for
the two cases.
Therefore the Drell-Yan processes 
are not sufficient
to distinguish between 
an asymmetric SU(2) sea and p-n
isospin symmetry breaking.

Detailed arguments can be found in Ref.~[1-5].

\end{document}